\documentclass{jetpl-en}
\twocolumn

\title{Proximity Effect and Spontaneous Vortex Phase
in Planar SF-Structures}

\rtitle{Proximity Effect in Planar SF-Structures}

\sodtitle{Proximity Effect and Spontaneous Vortex Phase in Planar
SF-Structures}

\author{V.\ V.\ Ryazanov\/\thanks{ryazanov@issp.ac.ru},
        V.\ A.\ Oboznov\, A.\ S.\ Prokofiev\, S.\ V.\ Dubonos*}

\rauthor{V.\ V.\ Ryazanov\, V.\ A.\ Oboznov\, A.\ S.\ Prokofiev\,
S.\ V.\ Dubonos}

\sodauthor{Ryazanov, Oboznov, Prokofiev, Dubonos}

\address{Institute of Solid-State Physics, Russian Academy of
Sciences,\\
Chernogolovka, Moscow region, 142432 Russia,\\
 *Institute of Microelectronic Technology and Ultra-High-Purity Materials,\\
 Russian Academy of Sciences,\\
 Chernogolovka, Moscow region, 142432 Russia}

\dates{December 4, 2002}{*}

\abstract{The proximity effect in SF structures was examined. It
is shown that, due to the oscillations of the induced
superconducting order parameter in a ferromagnet, the critical
temperature of an $SF$-bilayer becomes minimal when the thickness
of the ferromagnetic layer is close to a quarter of the period of
spatial oscillations. It is found that the spontaneous vortex
state arisen in the superconductor due to the proximity of the
magnetic domain structure of a ferromagnet brings about noticeable
magnetoresistive effects. © 2003 MAIK "Nauka/Interperiodica". }


\begin{document}

\maketitle

In recent years, considerable interest has been shown in metallic
multilayer systems with alternating magnetic and nonmagnetic
layers. The normal metal-ferromagnet structures ($NF$ systems)
exhibiting giant magnetoresistance have already found practical
use in computer technology ~\cite{GMR}. Promising elements based
on the multilayer superconductor-ferromagnet structures
($SF$-systems), such as the $FSF$ spin valve~\cite{Tagirov},
Josephson $SFS$ $\pi$-junction~\cite{Ryaz1}, and others, were also
been suggested and studied. The coexistence of superconductivity
and ferromagnetism is the problem of long- standing interest. The
antagonism of these two phenomena differing in spin ordering is a
cause for the strong suppression of superconductivity in the
contact area of the $S$ and $F$ materials~\cite{old}. The
appearance of oscillating sign-reversal order parameter in the $F$
layer near the $SF$ interface~\cite{Ryaz1,Buz82,Rad91,Kontos} is
another fundamental consequence of the proximity a ferromagnet and
a superconductor. Inspite of numerous theoretical works,
experimental studies of the $SF$ structures are in their infancy.
In particular, the influence of a real domain structure on the
indicated and other phenomena in the SF systems still remains to
be studied. In this work, three different types of SF structures
differing in size and geometry were prepared and studied: a
continuous thin-film strip of $SF$ bilayer, a macroscopic
superconducting $S-SF-S$ bridge (Notarys-Mercereau
bridge~\cite{Not_Mer}), and a one-dimensional chain of submicron
(mesoscopic) $S-SF-S$ bridges. The goal of this work was to
observe the following effects: (i) influence of the $F$-layer
thickness and the sign-reversal order parameter on the critical
temperature $T_{c,SF}$ of the $SF$-bilayer; (ii) appearance of a
spontaneous vortex state due to the proximity of the magnetic
domain structure of a ferromagnet; and (iii) appearance of
additional resistive contributions in the $S-SF-S$-bridge chains
caused by the injection of nonequilibrium quasiparticles from the
SF regions into a superconductor.

Experimental studies were carried out on the bilayer $Nb-Cu/Ni$
$SF$-structures, in which the $Cu_{0.43}Ni_{0.57}$ alloy films
with the Curie temperature $T_C\sim 150 K$ were used as a
ferromagnetic layer~\cite{Aarts}. Bottom superconducting Nb layer
with a thickness of $9-11$ $nm$ (close to the coherence length)
was sputtered by dc-magnetron. Top copper-nickel alloy film was
deposited in a single vacuum cycle by rf-diode sputtering. Weak
ferromagnetism of the $Cu/Ni$ alloy allowed us to retain the
superconductivity of the SF bilayer with $T_{c,SF}$ close to the
standard helium temperatures of $2-4 K$ and compare the obtained
results with the results of Josephson experiments~\cite{Ryaz1} on
the $Nb-Cu_{0.43}Ni_{0.57}-Nb$ sandwiches, in which a weak
ferromagnetism was necessary for the fabrication of continuous
homogeneous $F$-layers whose thickness would be comparable with
the pair-decay length $\xi_F$. Inasmuch as the pair-decay length
in the layers of classical ferromagnetic materials $(Co, Fe, Ni)$
is close to $1 nm$, the fabrication of thin-film Josephson $SFS$-
junctions using these metals is a challenge. The use of
ferromagnetic alloys with low Curie temperatures allowed us to
increase the pair-decay length by several orders of magnitude and
observe the transition of a Josephson $SFS$-junction to the
$\pi$-state upon a decrease in temperature~\cite{Ryaz1}.

Figure 1 shows the experimental geometry and the measured critical
temperature $T_{c,SF}$ of the bilayer $SF$-structures with a
superconducting niobium layer of thickness $11 nm$, close to the
coherence length in the thin-film niobium ($7-8$ $nm$), and
different thicknesses $d_F$ of a ferromagnetic layer. The
superconducting transition width was $\sim 0.3 K$. The curve shows
the $T_{c, SF}$ values corresponding to the onset of transition,
its middle part, and completion. It is seen that the dependence of
$T_{c, SF}$ on $d_F$ is nonmonotonic and has minimum at a
ferromagnet thickness of $5-8$ $nm$. Such a dependence was
predicted in~\cite{Rad91} and first observed for a bilayer $Nb/Gd$
structure in~\cite{jiang95}. This phenomenon is caused by the
appearance of superconducting electron pairs with the nonzero net
momentum in the presence of the exchange field that gives rise to
the specific LOFF-state with the inhomogeneous sign-reversal order
parameter, as was predicted in 1964 by Larkin and
Ovchinnikov~\cite{LO} and Fulde and Ferrel~\cite{FF}. The induced
superconductivity in a ferromagnet near the $SF$-interface proved
to be a quite realizable LOFF modification
~\cite{Buz82,Rad91,Kupr92}. It was shown in~\cite{aarts97,Lazar}
that the spatial oscillations of the order parameter in an
$SF$-bilayer with the thickness $d_F$ on the order of coherence
length $\xi_F$ in the ferromagnet give rise to the oscillations of
the $SF$-interface transparency, providing the simplest
explanation for the nonmonotonic dependence of $T_{c,SF}$ on
$d_F$. Simple considerations suggest that the lowest barrier at
the $SF$-interface (lowest $T_{c, SF}$ ) corresponds to the
thickness $d_F$ close to $1/4$ of the period $\lambda_{LOFF}$ of
spatial oscillations of the induced superconducting order
parameter in the $F$-layer~\cite{Fom1}. A comparison of the curve
in Fig. 1 with the results of a detailed theoretical analysis was
carried out in ~\cite{Fom1,Fom2}. We also had a chance of
comparing the period of spatial oscillations with the results
obtained in the experiments with the Josephson $SFS$ sandwiches,
in which the same composition of $Cu/Ni$ alloy was used as a
Josephson interlayer and the same sputtering technique was applied
(for the details of preparation and study of the Josephson $SFS$
junctions, see~\cite{Ryaz1}). The transition to the
$\pi$-state~\cite{Ryaz1,Buz82,Kupr92,Ryaz2}, in which the order
parameter has different signs on different banks of the SFS
sandwich, occurs for the ferromagnetic interlayer thickness close
to a half-period of spatial oscillations of the order parameter.
In the $Nb-Cu_{0.43}Ni_{0.57}-Nb$ sandwiches, we observed this
transition~\cite{Ryaz3} for the $F$-layers with a thickness of $15
nm$, i.e., twice the thickness corresponding to the minimal $T_{c,
SF}$ of the bilayer $SF$-structure, in agreement with the
theoretical estimates~\cite{Ryaz1,Fom1}.

To study the resistive processes in the current flow along the
$SF$-bilayer in more detail, planar $S-SF-S$ structures were
prepared (their different projections are shown in the insets in
Figs. 2, 3). The $SF$-bilayer was situated only in the central
section of superconducting bridge and formed by a ferromagnetic
strip, which completely spanned the superconducting bridge and
suppressed superconductivity in a square area of $10\times 10$
$\mu m$. To avoid the effects discussed in the preceding
paragraph, the $F$-layer thickness was taken to be large enough
($18 nm$) to suppress appreciably the $S$-layer and exclude the
formation of a barrier associated with the oscillations of
superconducting order parameter in the ferromagnet. The bridges
with above-mentioned sizes and the superconducting bridges with
$F$-islands of submicron size described in the last section of
this article were formed using electron-beam lithography. A two
step resistive junction obtained with a minimal transport current
of $0.5$ $\mu A$ is shown in Fig. 2a. The higher temperature step
with normal resistance $R_n$ of the structure corresponds to the
superconducting transition in the ferromagnet-free thin niobium
film. The transition at $T_{c, SF} = 3.6-3.8 K$ corresponds to the
resistive transition in the $SF$- bilayer with thickness $d_S = 9$
$nm$ and $d_F = 18$ $nm$. As it is seen in Fig. 2b, the
lower-temperature transition broadens sizably with an increase in
transport current. However, at the temperature $T^* = 2.6-2.65 K$
new sharp step arises, evidencing the abrupt dramatic increase in
the critical current of the $S-SF-S$ bridge below this
temperature. The bridge current-voltage characteristics (CVC's)
for different temperatures (in the absence of applied magnetic
field) are shown in Fig. 3. One can clearly see that the jumplike
increase in the critical current below $T^*$ is associated with a
cardinal change of the resistive mechanism in the bridge. In the
temperature range $T^* < T < T_{c,SF}$, the characteristics
exhibit constant differential resistance corresponding to the
magnetic-flux flow regime. The behavior below $T^*$ is typical for
long superconducting bridges, in which the dissipation is caused
by the sequential appearance of vortex slip lines incipient at the
bridge edges. The appearance of each line is displayed on the CVC
as a new slanted step, which was experimentally recorded using a
repeated current scan in the corresponding area. The unexpected,
at first glance, zero-field flux-flow regime at high temperatures
can easily be explained by the presence, in the superconductor, of
a spontaneous vortex phase associated with the stray magnetic
field in the domain walls of the ferromagnetic film. The
appearance of the spontaneous vortex phase was theoretically
discussed for "superconducting ferromagnets" and multilayer
$SF$-structures in~\cite{Sonin,Lyuks}.

The appearance of the vortex phase lines in the superconducting
layer near the domain walls of the ferromagnetic layer with in-
plane magnetic anisotropy is shown in the inset in Fig. 4. The
correlation between the flux-flow resistance and the number of
domains (domain walls) is confirmed by the magnetic field
measurements (Fig. 4). Magnetic field was applied parallel to the
bilayer plane. The observed symmetric (i.e., even with respect to
the field sign) behavior of $R( H )$ is caused by the direct
action of the field on the superconducting film. The curves also
show positive magnetoresistance peaks at the "magnetization
reversal" fields corresponding to the sample coercive field (the
hysteresis loop $M( H )$ is schematically drawn above the $R( H )$
curve; coercive fields were measured in the magnetic and Hall
experiments). In the step region on the $R( T )$ curve (Fig.2b), i.e., at
temperatures $T$ slightly above $T^*$ and currents $I\geq I_c$ ,
the magnetoresistance coefficient can be rather large and exceed
$100\%$.

We now discuss the conditions for the appearance of spontaneous
vortex phase in the $SF$-bilayer and the value of critical
temperature $T^*$ for transition to the "Meissner" phase. The
lower critical field for the penetration of a perpendicular
magnetic field into the film is determined by the effective
penetration depth $\lambda_{\perp}=\lambda^2/d_s$ and its
temperature dependence ($\lambda$ is the field penetration depth
into a thick film). Using the parameters of our film, one can
estimate $H_{c1}(0)\sim 10-20 G$. This is comparable with the
estimates of stray fields in the domain structure of our weak
ferromagnet, if one assumes that the domain wall width is on the
order of the domain size ($\sim 0.2-0.5\mu m$). Therefore, $T^*$
is the temperature for which the stray field becomes comparable
with $H_{c1}(T)$. Below this temperature the field of
ferromagnetic film does not pierce the superconducting film
through, and the flux-flow regime ceases. This model is
additionally confirmed by the fact that the constant differential
resistance disappears from the CVCs of $S-SF-S$ bridges with the
$F$-island sizes on the order of $0.2\times 0.5\mu m$. The
ferromagnetic islands with this area are, in fact, single-domain,
so that the stray field in the region of such domain is
appreciably weaker than the field produced by the domain wall.

We undertook attempt to connect the neighboring $SF$-regions in a
one-dimensional chain of $S-SF-S$ bridges with each other using
spin-polarized quasiparticles injected into the ferromagnet-free
sections of the superconducting film. As is illustrated in the
inset in Fig. 5, a section of the initial $SF$-bilayer was "cut"
at a length of $50\mu m$ so as to form $SF$-bridges separated by
the sections of the $Nb$-film. Since the length $L_F= 0.2\mu m$ of
the ferromagnetic island remained constant, the spacing between
the islands was varied by changing their number $N$ in the
structure. The results obtained for three structures with the
superconducting sections $L_S = 1,$ $0.5$, and $0.2$ $\mu m$ and,
correspondingly, the number of $SF$ islands $N = 30,$ $70$, and
$100$ are presented in Fig. 5. In all cases, the bridge width was
equal to $0.5$ $\mu m$. The resistive transition curves are given
in the $T/T_c^{Nb}$ coordinates, because the critical temperatures
of the free $Nb$-sections were slightly different due to the fact
that the instant of time the $Cu/Ni$-layer etching was completed
could not be controlled accurately, so that the niobium layers
were slightly different in thickness. In addition to this
transition and a rather smeared resistive transition for the
$SF$-islands, a new step, associated with the resistance caused by
the nonequilibrium quasiparticle injection into the
superconducting sections, evolves in the mid-transition region
starting at $L_S = 1$ $\mu m$. At $L_S = 0.2$ $\mu m$ (N=100),
this contribution becomes dominant. The estimate of the
penetration depth of nonequilibrium quasiparticles
(charge-imbalance relaxation length $\lambda_Q$) into
superconducting $Nb$ at temperatures $T_c^{Nb}$ close to gives a
value comparable to $0.2$ $\mu m$. For the case of spontaneous
antiparallel alignment of the magnetizations in the neighboring
$F$-islands, one could expect noticeable magnetoresistive effects
in a magnetic field applied perpendicular to the bridge chain in
the layer plane. Nevertheless, no appreciable effects were
observed, most probably, because of the absence of
antiferromagnetic alignment and the weak spin polarization of
quasiparticles in our system.

In summary, the
proximity effect in the $SF$-system have been studied in this
work; it is shown that, due to the spatial oscillations of the
induced superconducting order parameter in a ferromagnet, the
critical temperature $T_{c, SF}$ of a bilayer has a minimum when
the thickness of the ferromagnetic layer is close to a quarter of
the period of spatial oscillations. The occurrence of a
spontaneous vortex state caused by the proximity of the domain
magnetic structure of a ferromagnet has been observed in a
superconductor. In this state, the magnetoresistive effects are
quite appreciable.

We are grateful to A.I. Buzdin, A.A. Varlamov,
E.B. Sonin, and I.F. Lyuksyutov for helpful discussions and to
N.S. Stepakov for assistance in conducting the experiment. This
work was supported by the Russian Foundation for Basic Research,
the Ministry of Science of the Russian Federation, CRDF (grant no.
RP1-2413- CG- 02), and NATO (grant no. PST CLG 978153).

Fig. 1. Critical temperature of a bilayer $Nb-Cu_{0.43}Ni_{0.57}$
structure vs. the thickness of ferromagnetic layer.

Fig. 2. Resistive transition of the $S-SF-S$ bridge: (a) full
curve for a current of $0.5\mu A$ and (b) a part of the transition
corresponding to the $SF$ bilayer for currents $0.5$, $1$, $10$,
$30$, $80$, and $110\mu A$.

Fig. 3. Current-voltage characteristics of the $S-SF-S$ bridge at
temperatures $3.47$, $3.2$, $2.89$, $2.66$, $2.6$, and $2.49 K$.

Fig. 4. Resistance of the $S-SF-S$ bridge vs. longitudinal
magnetic field at $T = 2.66 K$ and a current of $50\mu A$. Arrows
indicate the field-scan direction. The magnetization curve for a
$Cu/Ni$ layer is schematically shown at the top.

Fig. 5. Resistive transitions of a one-dimensional chain of
$S-SF-S$ bridges: $N = (1)$: $30$, $(2)$: $70$, and $(3)$: $100$.

\vfill\eject
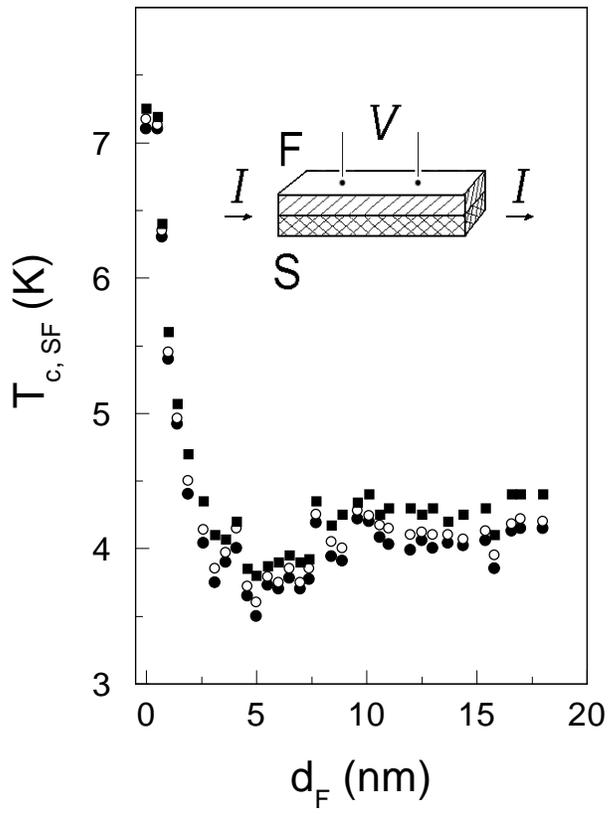
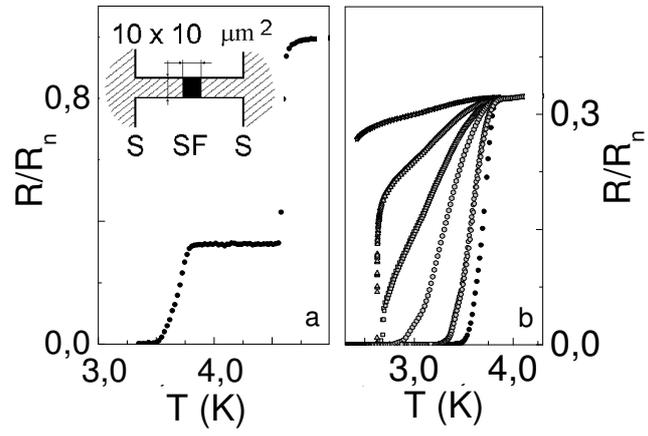
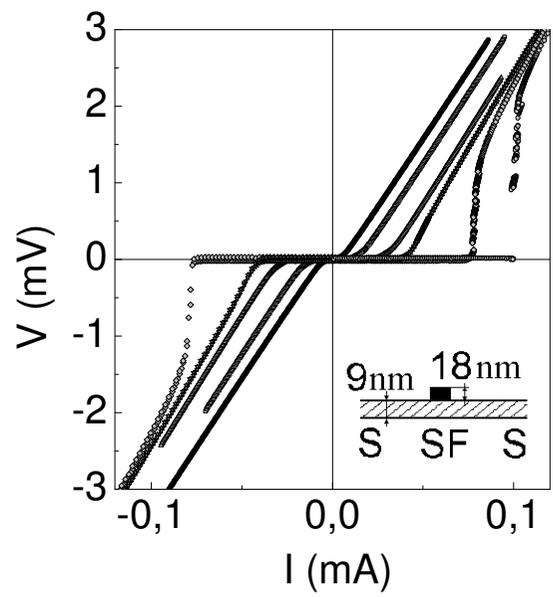
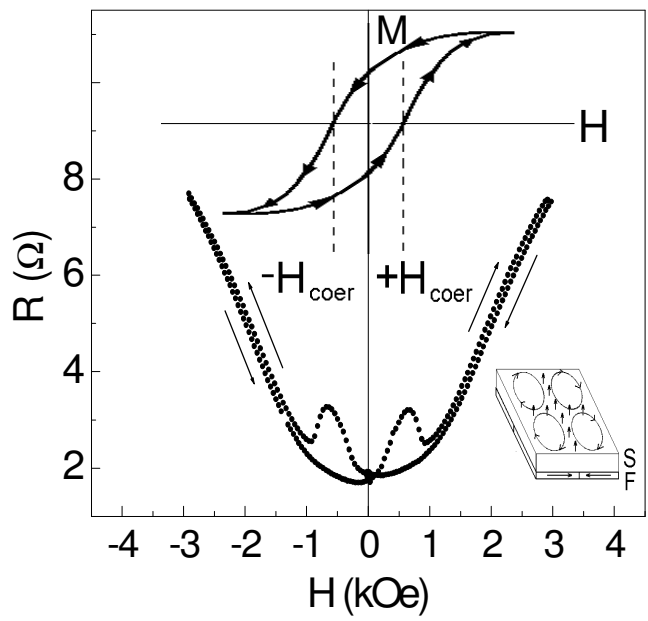
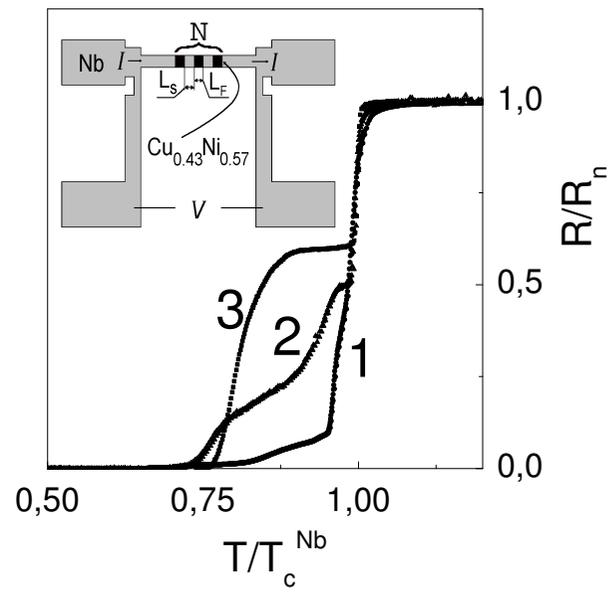
\end{document}